# Thermal conversion of ultrathin nickel hydroxide for wide bandgap 2D nickel oxides

Lu Ping[1*], Nicholas Russo[2*], Zifan Wang[3], Ching-Hsiang Yao[3], Kevin E. Smith[1,2†], Xi Ling[1,3,4†]

[1] Division of Materials Science and Engineering, Boston University, 15 St. Mary's Street, Boston, MA 02215, USA.

[2] Department of Physics, Boston University, 590 Commonwealth Avenue, Boston, MA 02215, USA.

[3] Department of Chemistry, Boston University, 590 Commonwealth Avenue, Boston, MA 02215, USA.

[4] The Photonics Center, Boston University, 8 St. Mary's Street, Boston, MA 02215, USA.

* These authors contribute equally to this work

†Corresponding author. Email: xiling@bu.edu, ksmith@bu.edu

**ABSTRACT**

Wide bandgap (WBG) semiconductors ($E_g$ >2.0 eV) are integral to the advancement of next generation electronics, optoelectronics, and power industries, owing to their capability for high temperature operation, high breakdown voltage and efficient light emission. Enhanced power efficiency and functional performance can be attained through miniaturization, specifically via the integration of device fabrication into two-dimensional (2D) structure enabled by WBG 2D semiconductors. However, as an essential subgroup of WBG semiconductors, 2D transition metal oxides (TMOs) remain largely underexplored in terms of physical properties and applications in 2D opto-electronic devices, primarily due to the scarcity of sufficiently large 2D crystals. Thus, our goal is to develop synthesis pathways for 2D TMOs possessing large crystal domain (e.g. >10 µm), expanding the 2D TMOs family and providing insights for future engineering of 2D TMOs. Here, we demonstrate the synthesis of WBG 2D nickel oxide (NiO) ($E_g$ > 2.7 eV) thermally converted from 2D nickel hydroxide (Ni(OH)$_2$) with the lateral domain size larger than 10 µm. Moreover, the conversion process is investigated using various microscopic techniques such as atomic force microscopy (AFM), Raman spectroscopy, transmission electron microscopy (TEM) and X-ray photoelectron spectroscopy (XPS), providing significant insights on the morphology and structure variation under different oxidative conditions. The electronic structure of the converted Ni$_x$O$_y$ is further investigated using multiple soft X-ray spectroscopies, such as X-ray absorption (XAS) and emission spectroscopies (XES).

**INTRODUCTION**

WBG semiconductors, with a bandgap in the range of 2.0-4.0 eV, have demonstrated great potential in the field of electronics, optoelectronics, and power industries due to their power handling capability and high temperature resilience.[1] Specifically, direct mapping of the electric field in the device channel of gallium nitride (GaN) based high-electron-mobility transistors (HEMTs) is achieved with a sub-micrometer resolution, allowing better understanding of how electronic devices work and their potential limitations.[2] Besides, development of ultraviolet photodetectors are flourished by the exploration on various WBG semiconductors and hybrid structures, e.g., metal (di)-chalcogenides,[3] perovskite,[4] owing to their direct light emission in the ultraviolet range. Moreover, WBG semiconductors are also widely considered as the current state-of-the-art and future answer in power electronics like smart grid applications.[5] To further improve the efficiency and lower the cost, miniaturized electronic and optoelectronic devices constructed and accomplished by 2D semiconductors are believed to be the future for next generation micro-nano electronics.[6–11] So far, over a thousand of possible inorganic compounds have been discovered as layered materials,[12] with a high concentration on 2D narrow bandgap (NBG) semiconductors ($E_g$ < 2.0 eV).[13–18] In contrast, the research and development on WBG 2D semiconductors (Eg > 2.0 eV) consists of metal chalcogenides, halide system and metal oxides (MOs) still lags behind that of conventional NBG semiconductors.[12,19,20] However, it is imperative that equal importance and emphasis are placed on advancing the understanding and applications of WBG 2D semiconductors to drive the forthcoming revolution in 2D micro-nano devices.

TMOs, of which the bandgaps are mostly located in the range of 3.0–5.0 eV, is complementary to NBG semiconductors.[21] As an essential subset of WBG semiconductors, their bulk form are widely applied in electronic and optoelectronic fields benefiting from their visible-light transparency, low cost and high stability in environments. However, the 2D form of TMOs is far less explored, although they may inherit the attracting properties of their bulk form or exhibit unprecedented properties due to quantum confinement effects and interfacial interactions.[21] Recent research has realized the preparation of several atomically thin 2D TMOs through different approaches such as physical/chemical assisted exfoliation,[22,23] liquid metal oxidation,[24,25] and vapor phase deposition (e.g., physical or chemical vapor deposition and atomic layer deposition).[26,27] The obtained 2D TMOs have substantially expanded the scope of the 2D family for next-generation opto-electronics. For example, the recently developed novel p-type 2D hexagonal $TiO_2$ (h-$TiO_2$) demonstrates a high hole mobility up to 950 $cm^2\,V^{-1}\,s^{-1}$ in back-gated field-effect transistors (FETs);[22] photodetectors based on 2D $Fe_2O_3$ realize ultrabroadband response ranging from UV (375 nm) to long wavelength infra-red (IR) (10.6 µm).[28] Nevertheless, many other 2D TMOs such as $Ni_xO_y$, which have demonstrated high hole mobility and fast switching speed in nano FETs,[29] are not widely realized yet. Especially, the ones with sufficiently large lateral domains (>10 µm) are essential for semiconductor device design and application. Therefore, it is highly desirable to develop synthesis strategies to fulfill this objective, paving the way for both fundamental study and application exploration of WBG 2D semiconductors.

In this work, we realized the synthesis of WBG 2D (>2.7 eV) NiO through a simple thermal conversion from 2D $Ni(OH)_2$. The morphology of the 2D NiO crystals closely resembles that of the precursor 2D $Ni(OH)_2$, with lateral dimensions exceeding 10 µm, as previously reported in our research.[30] Specifically, the synthesis of 2D NiO involved air oxidation of 2D α-$Ni(OH)_2$ crystals under varying oxidizing temperatures, allowing for the establishment of structural and chemical composition changes throughout the thermal conversion process at different temperature regimes. The electronic bandgaps of the resulting $Ni_xO_y$ ($E_g$ > 2.7 eV) are determined using soft X-ray techniques, i.e., X-ray absorption spectroscopy (XAS) and X-ray emission spectroscopy (XES). Comparative analysis of the obtained bandgap values with those of other reported WBG 2D semiconductors verified that it sits firmly in the middle range of MOs and at the higher end of metal chalcogenides and halide systems. We anticipate the insights gained from this study and the characteristics of the synthesized 2D $Ni_xO_y$ to offer valuable guidance for engineering 2D MOs and designing applications involving WBG 2D semiconductors.

**RESULTS AND DISCUSSION**

2D α-$Ni(OH)_2$ flakes are deposited via drop casting on 1×1 cm $SiO_2$/Si substrates, and the samples undergo oxidation in air, as illustrated in the schematic presented in **Fig. 1**. Following a 10-hour oxidation process at 100 °C, the resulting sample is labeled as NiO-100 and is subjected to characterizations using various techniques. Subsequently, the sample is subjected to heating at 200 °C and maintained for 10 hours, leading to NiO-200, which is characterized before undergoing further oxidation at 300 °C for an additional 10 hours. The highest temperature explored and conducted in this work is 1000 °C, resulting in NiO-1000. Throughout each oxidation step, ramping and cooling rates are set at 100 °C per hour, ensuring a steady and smooth conversion process. Multiple characterizations such as Raman spectroscopy, and XPS are performed at each step, showing 2D NiO starts to form at 300 °C.

**Morphology characterization of 2D NiO**

**Fig. 2A** is the optical image of an area that is covered by 2D α-$Ni(OH)_2$ flakes with three flakes are contoured by red, blue and purple dash lines. **Fig. 2B** is the optical image of the same area after converting it to 2D NiO-300, showing the three contoured flakes stay in their original shapes without visible decomposition. AFM is applied to track the thickness variation along the increased oxidizing temperature. As shown in **Fig. 2C**, the thickness of this 2D α-$Ni(OH)_2$ flake is measured as 15.8 nm, at the same position, the thickness of NiO-300 decreases dramatically by 63.3% to 5.8 nm (**Fig. 2D**), which attributes to the removal of interlayer intercalation and the crystal structure change from hydroxide, a vdW (layered) to non vdW oxide (close-packed). The thickness does not change significantly when the temperature further increases, it slightly decreases to 4.3 nm on NiO-1000 (**Fig. 2E**). In general, the dramatic vertical reduction along the conversion occurs at 300 °C, as evidenced by the tracked thickness variation (Fig. S1-5) observed on five 2D α-Ni(OH)$_2$ flakes (**Fig. 2F**). Meanwhile, the similar phenomenon is also observed in thermogravimetric analysis (TGA), which is applied to track the weight change with the increasing temperature from the starting 2D α-Ni(OH)$_2$ to 500 °C. Dramatic weight losses are observed (Fig. S6) at 100 and 300 °C due to the removal of surface water and intercalation.

**Structural characterization of 2D NiO**

First, Raman spectroscopy is applied to identify NiO converted from Ni(OH)$_2$ since it provides unique vibrational information and fingerprints of different materials. A deep UV laser with 266 nm wavelength (~ 4.66 eV) is used to excite the resonance. As shown in **Fig. 3A**, the characteristics peaks (red arrows) from 2D α-Ni(OH)$_2$ at 502.8 and 809.9 cm$^{-1}$ are diminishing when the temperature is increased to 300 °C, which completely disappear when temperature reaches 800 °C. This suggests the 2D α-Ni(OH)$_2$ is completely consumed and converted to a new substance. Meanwhile, at 300 °C, a new peak at 1087.2 cm$^{-1}$ rises and expands to a slightly wider peak at 600 °C, which eventually becomes distinct and significant at 800 °C and corresponds to the 2LO (longitudinal optical) mode of NiO.[31] Besides, LO, 1M (magnon) and 2M modes of NiO are also observed at 567.9, 718.9 and 1437.1 cm$^{-1}$ at 800 °C,[31,32] suggesting the conversion is complete at this temperature. All the characteristic peaks of NiO resident at the same position when the temperature is lifted to 1000 °C. Therefore, the results show the conversion of 2D α-Ni(OH)$_2$ to 2D NiO starts at 300 °C and completes at 800 °C.

Moreover, the emergence of NiO is observed by TEM as well. As shown in **Fig. 3B**, the low magnification TEM reveals the 2D nature of a micrometer large domain of a typical α-Ni(OH)$_2$ flake. High magnification TEM in **Fig. 3C** shows the d-spacings between (200) and (400) planes are 2.2 and 4.4 Å, respectively.[33] The SAED (**Fig. 3C inset**) provides sharp diffraction pattern and reveals the hexagonal lattice structure of Ni(OH)$_2$. Diffraction from (110), (2$\bar{1}$0) and (1$\bar{2}$0) lattice planes can be observed. While on NiO-300 (**Fig. 3D inset**), new diffraction pattern is observed within the original pattern and the d-spacings are measured as 2.4 and 2.1 Å, corresponding to (111) and (200) planes from NiO.[34] The spacing of these two planes can also be measured from high magnification TEM shown in **Fig. 3D**, which is 2.4 and 2.2 Å for (111) and (200), respectively.[34]

**Chemical states variation during the conversion**

After identifying the starting and ending points of the conversion from hydroxide to oxide, we continue to investigate the entire process using XPS by charactering the chemical states of Ni and O that occurred at different oxidizing temperatures during the conversion. Firstly, Ni 2p spectra is collected along with the increasing oxidizing temperature and is shown in **Fig. 4A** (separate spectrum for each sample can be found in Fig. S7). In general, the most noticeable characteristic is at 300 °C, where a new peak rises up that denotes to NiO. Specifically, Fig. S7A shows the Ni 2p spectrum of the original 2D α-Ni(OH)$_2$, and a spin-orbital coupling caused splitting energy between Ni 2p$_{3/2}$ (855.9 eV) and Ni 2p$_{1/2}$ (873.4 eV) is measured as 17.5 eV, which is characteristic of the Ni$^{2+}$ ion in Ni(OH)$_2$.[35,36] For NiO-300 (**Fig. 4A**), the Ni 2p$_{3/2}$ (blue arrow) and Ni 2p$_{1/2}$ (orange arrow) peaks both are comprised of two peaks, of which the binding energy positions suggest the formation of NiO. Besides, there is no distinct change when the temperature is further lifted to 800 °C (Fig. S7I) and 1000 °C (Fig. S7K), suggesting that there is no new substance formed in the higher temperature range.

Meanwhile, O 1s spectra is collected along with the increasing oxidizing temperature and is shown in Fig. S8. A new peak at 528.6 eV also rises up at 300 °C, which is highly consistent with our observation from Ni 2p analysis and previous structure characterizations. Specifically, Fig. S9A shows the O 1s spectrum of the original 2D α-Ni(OH)$_2$, which consists of three peaks at 531.1, 532.4 and 533.2 eV, representing Ni-O-H, O-Si-O (from the substrate) and H-O-H, respectively.[37,38] The appearance of H-O-H signal suggests the existence of H$_2$O in the crystals, mainly as intercalation species between layers to form α-Ni(OH)$_2$. After the sample is oxidized under 300 °C (Fig. S9D), the peak representing water completely disappears, suggesting the removal of the intercalated water, which is consistent with the dramatic thickness decrease from AFM measurements (**Fig. 2F**). Besides, a new peak at 528.6 eV shows up that is denoted as NiO, and the peak at 530.4 eV suggests the formation of Ni$_2$O$_3$.[39,40] With the increasing oxidizing temperature, the peak representing Ni$_2$O$_3$ diminishes at 800 °C (Fig. S9I) and eventually almost disappears at 1000 °C (Fig. S9K) due to the thermal decomposition from Ni$_2$O$_3$ to NiO.[41,42]

To prove our hypothesis of the formation and decomposition of Ni$_2$O$_3$, the peak areas in XPS across various temperatures are analyzed and compared to investigate the thermal conversion process. As shown in **Fig. 4B**, peak area in dark blue (area A) is solely contributed to O$^{2-}$ from NiO. However, the peak area in red (area B) is more complex, it could be the contribution of O$^{2-}$ from Ni(OH)$_2$ or from Ni$_2$O$_3$. Similarly, the peak area on Ni 2p$_{3/2}$ spectra is colored as well and shown in **Fig. 4C**. Peak area in light blue (area A) is contributed to Ni$^{2+}$ from NiO while the peak area in red (area B) could be the contribution of Ni$^{2+}$ from Ni(OH)$_2$ or Ni$^{3+}$ from Ni$_2$O$_3$, of which the Ni ions have very close binding energy thus very hard to distinguish one from the other.[38,43] Area B may also include the secondary electrons from area A, which is an inherent characteristic of XPS.[44] Therefore, the peak area ratio B/A is discussed and analyzed (**Fig. 4D**) instead of the exact peak area to track the relative amount of NiO formed during the conversion and to investigate the chemical reactions occurred under different temperature range. In general, the B/A values obtained from Ni $2p_{3/2}$ and O 1s show similar trends, which increase first in lower temperature range (below 500 °C, **stage I**), then dramatically decrease in medium temperature range (500 to 800 °C, **stage II**) and eventually stabilize in higher temperature range (above 800 °C, **stage III**). During **stage I**, formation of both NiO and $Ni_2O_3$ happen, which increases B and A values simultaneously, resulting in the overall B/A value increase. When the temperature continues getting higher (**stage II**), $Ni_2O_3$ starts to decompose and generates NiO and $O_2$ due to its poor thermal stability at high temperature (~ 600 °C),[42] leading to a lower B and higher A, thus gives an overall decreasing B/A value. After the $Ni_2O_3$ decomposition is complete (**stage III**), the B/A value stabilizes, the generated $O_2$ leaves the material and creates oxygen vacancies ($V_{oxy}$) in the crystals.

**Electron structure characterization of 2D NiO**

The conversion process from the initial 2D α-Ni(OH)$_2$ to NiO is then tracked using XAS at the O K edge (**Fig. 5A** orange) and the Ni L edge (Fig. S10). Combined with non-resonant XES at the O K (**Fig. 5A** blue), the bandgap of each sample annealed at different temperatures is estimated.[45,46] As can be seen from **Fig. 5A**, the O K edge changes in structure, with the peak shifts from 533.5 eV for α-Ni(OH)$_2$ (purple arrow) to 532.5 eV in NiO-300 (green arrow). Higher temperature annealing results in sharper features appearing at higher energies, which is attributed to the healing of defects as the NiO sample approaches its bulk counterpart.[47] The Ni L edge shown in Fig. S10, does not show evidence of strong sensitivity to whether the material is in the Ni(OH)$_2$ or NiO phase, which we attribute to the lack of change of symmetry and charge at the $Ni^{2+}$ site.[48]

The band edges are determined using the second derivative method (**Fig. 5A** dash lines),[49] where the onset (short orange lines for absorption and short blue lines for emission) of the conduction (valence) band is determined by the first peak in the rising spectral edge above the noise level, corresponding to the change in the concavity of the onset absorption (emission) spectrum.[50–52] To smooth the spectra to make estimation of the second derivative feasible, we employ a Savinzky-Golay filter using a window size of 15 datapoints and polynomial of orders 3 and 4 for XES and XAS data, respectively.[53] **Fig. 5B** shows the absorption (orange dots) and emission (blue dots) band edges position derived from **Fig. 5A**, and the bandgap is estimated by taking the difference between absorption and emission onsets for each sample, as shown in **Fig. 5C**. Notably, to account for the core-hole final state effect in the XAS spectra, 0.3 eV is added to the core-hole screening for O K edge.[54,55] As can be seen in **Fig. 5A**, the bandgap of the $Ni(OH)_2$ film is determined to be around 2.2 eV, which is significantly lower than the bulk value. Such exceptional deviation from the quantum size effect of nanoscale systems has been predicted from DFT calculation[56] and also observed in several other systems such as hematite and zinc nanoparticles.[57,58]

The reason behind the deviation could be the low lying unoccupied surface states in the conduction band arising from O-H surface bonds. A similar effect has been reported in 6 nm FeOOH nanoparticles where the bandgap is reduced from the bulk counterpart, where the reduction in gap is attributed to structural disorder and under coordinated oxygen sites at the particle surface.[59] In both FeOOH nanoparticles and 2D $Ni(OH)_2$ cases, surface states play a significant role in determining the bandgap. And a previous dual computational and experimental study of NiO (111) and (100) surfaces has demonstrated that hydroxyl desorption begins at 126 °C.[60] If the lower bandgap in $Ni(OH)_2$ is due to O-H, then the removal of these states should increase the gap; this appears consistent with our XAS and XES data where the determined bandgap increases to 2.9 eV in the 100°C annealed sample (**Fig. 5C**).

As the sample is annealed above 100 °C, the bulk of the sample starts to convert to NiO by the removal of $H_2O$ molecules from the bulk, resulting in the formation of vacancies, while under atmospheric conditions the sample is re-oxidized by ambient $O_2$. The 2D NiO sample with highest bandgap is from annealed at 1000 °C and the determined gap is 2.75 ± 0.24 eV, while the values for the other 2D NiO samples hover between 2.0 to 2.72 eV . Other studies have also found band gaps in nanostructured NiO to be less than the bulk value.[61] In comparing the bandgap of 2D NiO with other reported WBG 2D oxides (**Fig. 5D**),[62,63] we determine the gap sit firmly in the middle range of MOs and at the higher end of metal chalcogenides and halide systems (**Fig. 5E**).[20,62]

**CONCLUSION**

In conclusion, we achieve the largest ever reported 2D NiO (>10 μm) down to a few nanometers though a thermal conversion of 2D $Ni(OH)_2$, with a comprehensive understanding of the process. This conversion from a vdW to a non vdW structure not only sets a precedent for producing more 2D non vdW materials but also unveils insights into morphology, structural variations, and chemical reactions during the process, offering valuable guidance for material design, integration and manufacturing. Moreover, the bandgap of the thermally converted 2D $Ni_xO_y$ ($E_g$ > 2.7 eV) is observed higher than that of the starting material 2D α-$Ni(OH)_2$, affirming the position of 2D NiO within the WBG 2D semiconductors family. The electronic band structure of $Ni_xO_y$ ($E_g$ > 2.7 eV) examined by soft X-ray spectroscopy, XAS and XES, contributes valuable knowledge on the understanding of this material type. Considering the simplicity, great reproducibility and industrial fabrication process compatibility of the thermal conversion method, our work demonstrates a promising pathway for producing non vdW 2D materials from layered precursors, thereby expanding the 2D materials family. And the pivotal results on 2D NiO suggests its potential as a WBG semiconductor for high voltage operation in future micro-nanoelectronics.

**METHODS**

**Thermal Conversion.** The thermal conversion takes place in a mostly enclosed oven from Thermo Scientific (Lindberg/Blue M Moldatherm Box Furnace, 5.3L), in which the air can only exchange with the reservoir via a 10 cm-diameter hole.

**Chemicals and Materials.** 2D Ni(OH)$_2$ flakes used for thermal conversion are from our previous work.[30]

**Characterization Techniques.** Optical images are collected by optical microscopy (Nikon DS-Ri2). AFM topography is acquired on a Bruker Dimension 3000 in a tapping mode. TGA is measured on a SDT Q600 from TA Instruments. XPS spectra are collected from Thermo Scientific Nexsa. TEM measurements are performed on a FEI Tecnai Osiris TEM, operating at a 200 keV accelerating voltage. SAED is also measured on a FEI Tecnai Osiris TEM. For the deep UV Raman, a 266 nm (4.66 eV) laser excitation is used and the spectrum is collected from a custom-built micro-Raman setup. The laser beam is directed onto the sample through a 40X ultraviolet objective lens (Thorlabs, LMU-40X-UVB), with a laser spot size of 1 μm. The Raman signal is collected using the same objective lens, focused into a 230 μm core fiber, and transmitted to the spectrometer (Jobin-Yvan, FNR640) equipped with a TE-cooled CCD (Andor, Newton).[64] Soft x-ray absorption and emission spectroscopies (XAS, XES) are carried out at the Advanced Light Source (ALS) at Lawrence Berkeley National Laboratory on beamline 8.0.1. using the iRIXS endstation. XAS is performed at the O K edge and Ni L edge and XES measurements, (excitation energy = 600 eV) is measured at the O K edge.[65]

**ACKNOWLEDGEMENT**

**Finance:** Acknowledgement is made to the donors of the American Chemical Society Petroleum Research Fund under Grant No. (PRF# 61965-ND10) for support of this research. L.P., N.R., K.S. and X.L. acknowledge the support of National Science Foundation (NSF) under Grant No. 2216008. Work done by X.L. and Z.F.W. is also supported by the U.S. Department of Energy (DOE), Office of Science, Basic Energy Science (BES) under Award Number DE-SC0021064 and the NSF under Grant No. 1945364. We thank Professor Yu-Ming Chang from National Taiwan University for using his deep-ultraviolet Raman system for a portion of this work.

**Facility:** This research uses resource from Center for Nanoscale Systems, Harvard University; Center of Condensed Matter Sciences, National Taiwan University. This research also uses resource from Advanced Light Source, which is a DOE Office of Science User Facility under contract no. DE-AC02-05CH11231. We acknowledge Dr. Haoxuan Yan for helping with the TGA measurement.

## AUTHORS CONTRIBUTION

L. P., N. R., X. L. conceived and designed the experiments. L. P. performed all the synthesis. L. P., N. R., Z .W. and C. Y. characterized the samples. L. P., N. R., K. S. and X. L. analyzed the data and wrote the manuscript. All authors discussed the results and commented on the manuscript.

## DATA AVAILABILITY

The data that support the findings of this study are available from the supplementary information.

## ADDITIONAL INFORMATION

The authors declare no competing interest.

**FIGURES**

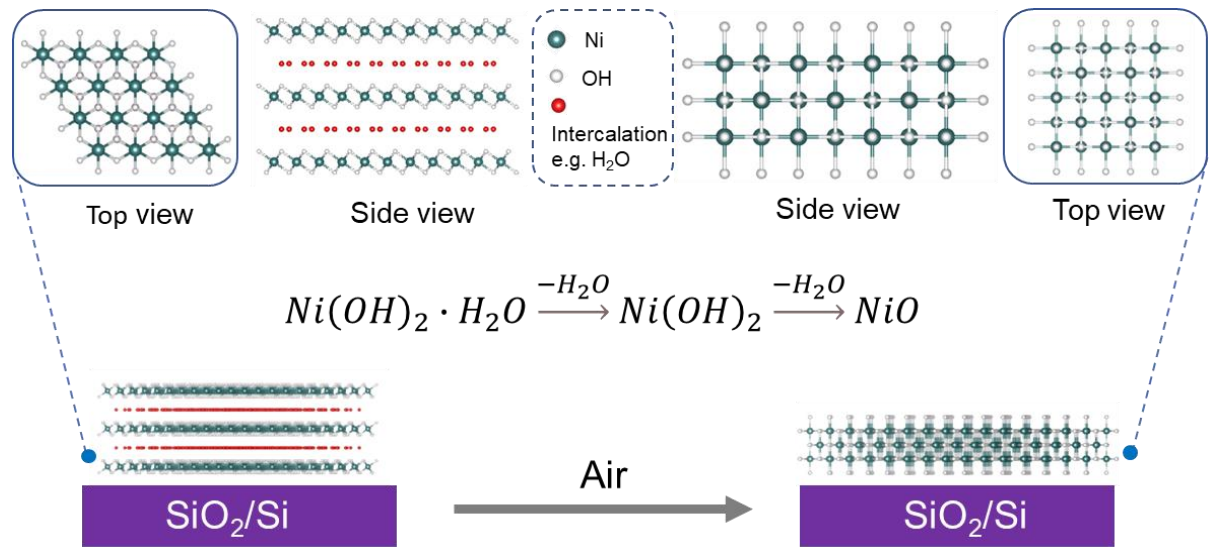

**Fig. 1.** Schematic of thermal conversion from 2D α-Ni(OH)$_2$ to 2D NiO.

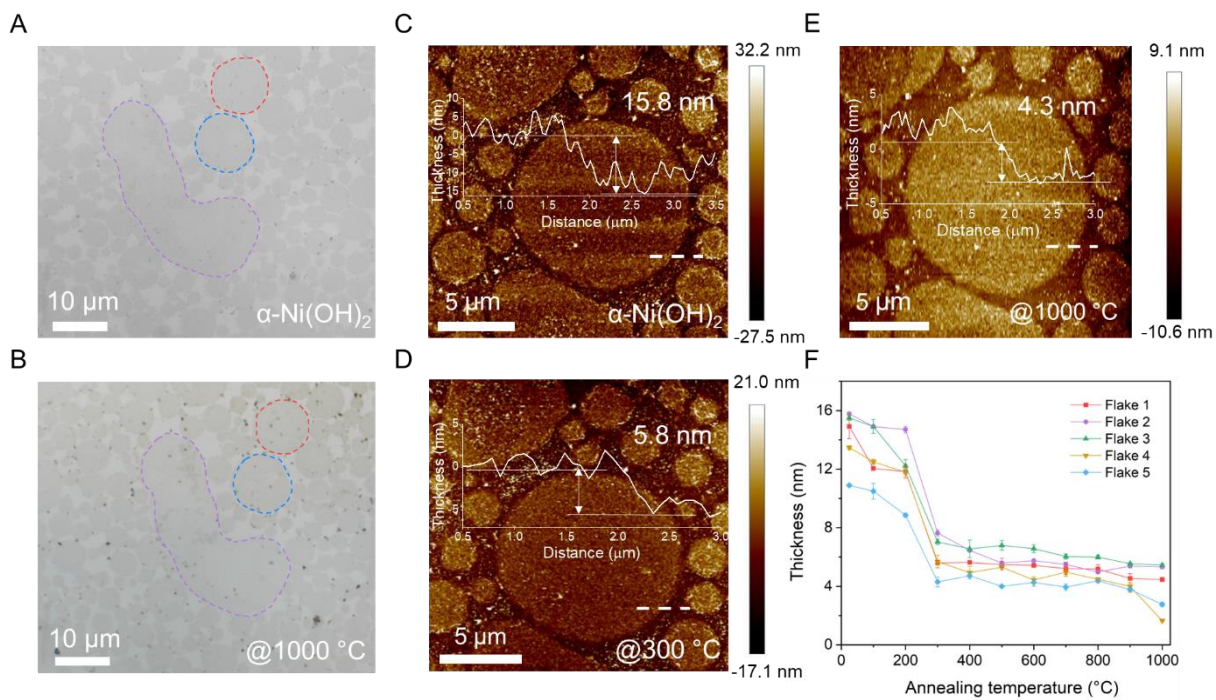

**Fig. 2. Morphology characterization of 2D NiO converted from 2D α-Ni(OH)₂.** Optical images of the 2D α-Ni(OH)₂ sample before (**A**) and after 10 hours oxidizing at 1000 °C (**B**). AFM images of the flake before (**C**) and after 10 hours oxidizing at 300 °C (**D**), 1000 °C (**E**). Insets: height profiles obtained along white dash lines. (**F**) Thickness variation with increased oxidizing temperature on seven 2D α-Ni(OH)₂ flakes.

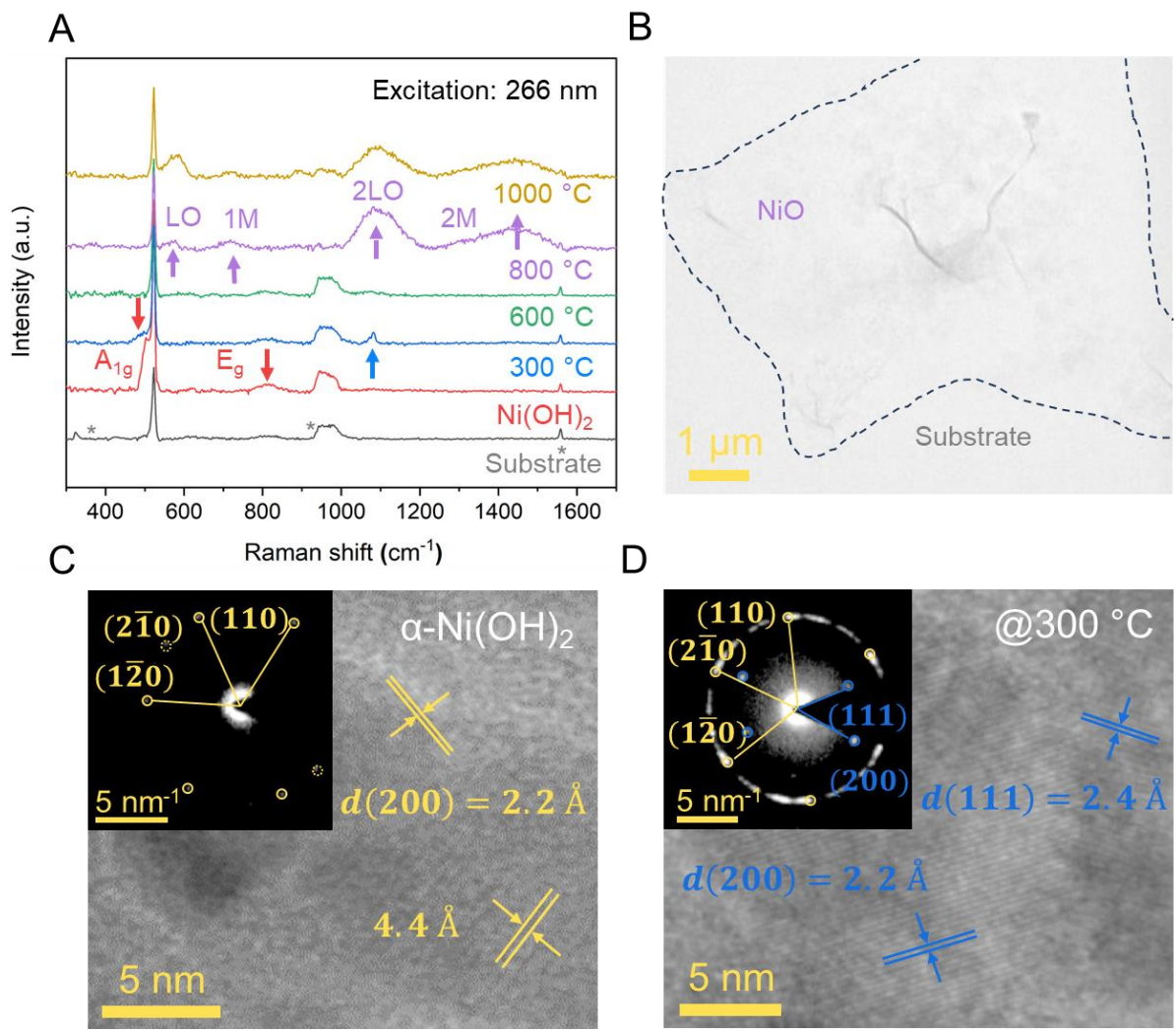

**Fig. 3. Identification of NiO and structure characterization.** (**A**) Deep UV Raman spectra of 2D α-Ni(OH)$_2$, NiO-300, 600, 800, 1000°C and the substrate excited with 266 nm laser. Peaks labeled with "*" are from the Si substrate. (**B**) Low-magnification TEM image of 2D α-Ni(OH)$_2$, showing the crystal domain size is over 10 µm. (**C**) High resolution TEM image of 2D α-Ni(OH)$_2$. Inset: the SAED pattern. (**D**) High resolution TEM image of NiO-300. Inset: the SAED pattern.

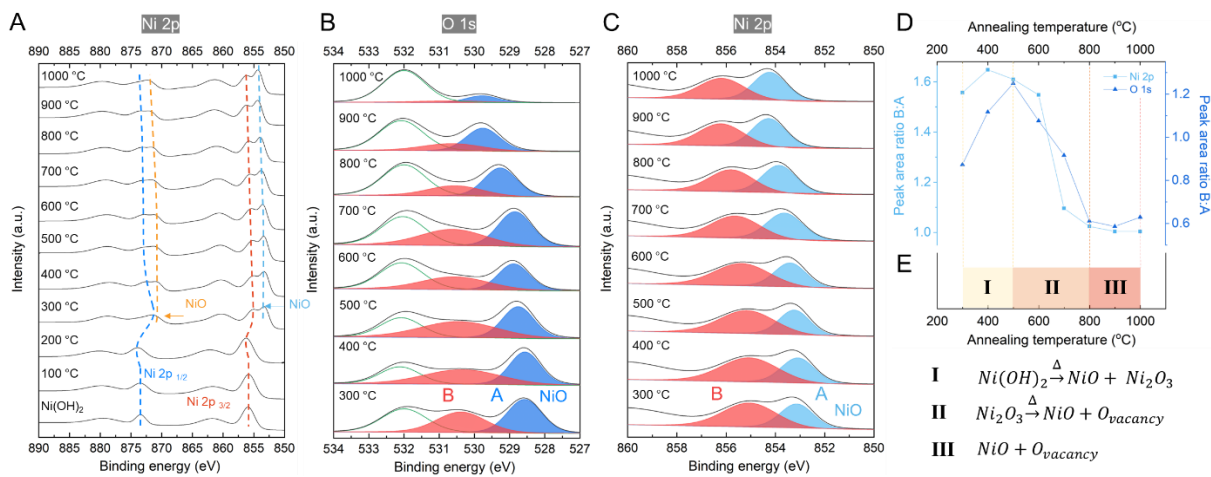

**Fig. 4. Investigation of the conversion process by XPS.** (**A**) Ni 2p spectra of NiO oxidized under different temperatures, from 100 to 1000 °C. (**B**) O 1s spectra of NiO oxidized under different temperatures, from 300 to 1000 °C (Whole range from RT to 1000 °C is in shown in Fig. S8). The colored area A in dark blue consists of contribution from NiO, the colored area B in red consists of contribution from both Ni(OH)$_2$ and Ni$_2$O$_3$. (**C**) Ni 2p$_{3/2}$ spectra from 300 to 1000 °C. The colored area A in light blue consists of contribution from NiO, the colored area B in red consists of contribution from both Ni(OH)$_2$ and Ni$_2$O$_3$. (**D**) Comparison of B/A from O 1s and from Ni 2p$_{3/2}$, which showing a similar trend. (**E**) The concluded chemical reactions during thermal conversion in three phases, I, II and III.

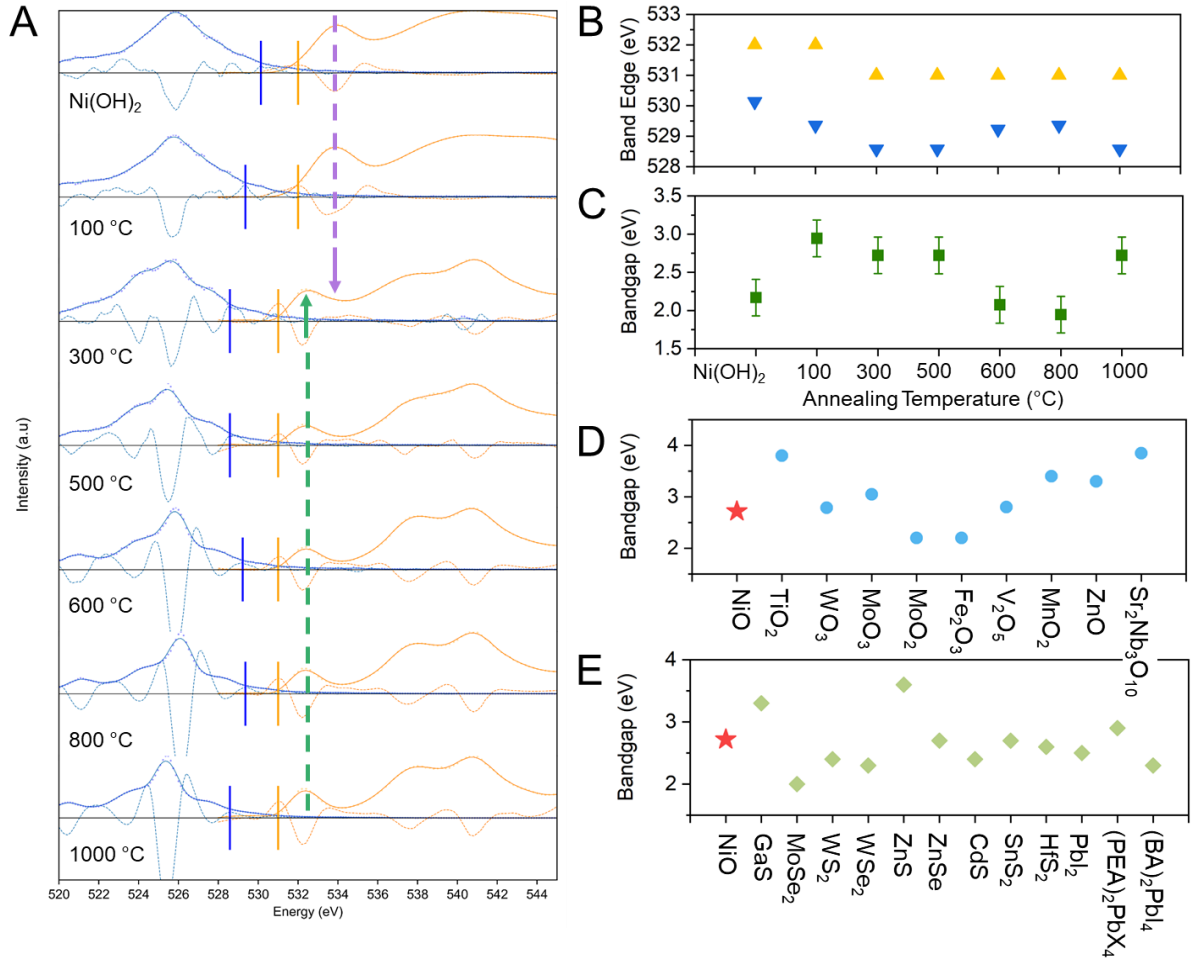

**Fig. 5. Electronic structure characterization of 2D Ni(OH)$_2$ and NiO films using soft X-ray techniques.** (**A**) XAS (TEY, Total Electron Yield) (orange) and XES (blue) spectra in solid lines. Savinzky-Golay filter is applied to reduce noise and compute second derivatives (dashed lines, magnified), band onsets (short orange lines for absorption and short blue lines for emission) are determined by the first peak in the rising spectral edge above the noise level. Purple and green arrows indicates the structure change at 533.5 eV of α-Ni(OH)$_2$ and at 532.5 eV in NiO, respectively. Spectra are normalized to highest intensity peak in each spectrum. (**B**) Band edges derived from absorption (orange dots) and emission (blue dots) spectra. (**C**) The determined bandgaps, by taking the difference between absorption and emission onsets from (**C**) for each sample. (**D**) Bandgap of 2D NiO (in red), in comparison to other WBG 2D oxides.[62,63] (**E**) Bandgap of 2D NiO, in comparison to other WBG 2D metal chalcogenides and halide systems.[20,62]